\documentclass[a4paper,11pt]{article}
\usepackage{pos}
\usepackage{slashed}
\usepackage{subfig}
\usepackage{booktabs}
\newcommand{\FI}{\mathcal{I}}
\newcommand{\TOP}[1][top]{\mathsf{#1}}
\newcommand{\PL}{\TOP[PL]}

\newcommand{\NA}{\TOP[NA]}
\newcommand{\NB}{\TOP[NB]}

\newcommand{\FIPL}[1]{\FI_{\PL}(#1)}

\newcommand{\FINA}[1]{\FI_{\NA}(#1)}
\newcommand{\FINB}[1]{\FI_{\NB}(#1)}

\newcommand{\HyperInt}{\href{http://bitbucket.org/PanzerErik/hyperint/}{\texttt{\textup{HyperInt}}}}
\newcommand{\td}{\mathrm{d}}
\newcommand{\Li}[1]{\operatorname{Li}_{#1}}

\title{Next-to-leading order mixed QCD-electroweak corrections to Higgs production at the LHC}
\ShortTitle{NLO QCD-EW $PP \to H$}

\author*[a]{Marco Bonetti}

\affiliation[a]{Institute for Theoretical Particle Physics and Cosmology, RWTH Aachen University, \\Sommerfeldstrasse 16, D-52056 Aachen, Germany}

\emailAdd{bonetti@physik.rwth-aachen.de}

\abstract{After ten years from its discovery, the Higgs boson is still under unprecedented scrutiny. A huge theoretical effort has been invested in modelling Higgs boson production through gluon fusion, reaching $\text{N}^3\text{LO}$ predictions in pure QCD. This incredible theoretical achievement makes the exact computation of sub-leading contributions, such as mixed QCD-Electroweak corrections, necessary. I will present the analytic calculation of the gluon- and quark-initiated two-loop four-point contributions to such class of corrections mediated by light quarks at order $v \alpha^2 \alpha_S^{3/2}$.

~

TTK-22-26,~P3H-22-081}

\FullConference{%
  41st International Conference on High Energy physics - ICHEP2022\\
  6-13 July, 2022\\
  Bologna, Italy
}

\begin{document}
\maketitle

\section{Motivations}
\label{sec:intro}

Even after ten years from its joint discovery at ATLAS \cite{Aad:2012tfa} and CMS \cite{Chatrchyan:2012ufa}, the Higgs boson remains under unprecedented scrutiny. After two LHC runs the amount and quality of data allowed for extremely precise and accurate experimental results on many production and decay channels for Higgs boson production \cite{ATLAS:2022tnm,CMS:2022uhn}, and many projections for the High-Luminosity phase of the LHC foresee a decrease of the total error associated to many Higgs production modes to decrease significantly. In particular, Higgs production through gluon fusion, the main channel of production for the Higgs boson at the LHC, is expected to reach error of the order of the percent, or even lower. Such remarkable results must be met by equally reliable theoretical predictions.

The theory community has undergone a long and challenging journey to produce theoretical predictions with percent-level uncertainties, leading to astonishing results such as the evaluation of $\text{N}^3\text{LO}$ QCD corrections for Higgs boson production in gluon fusion \cite{Anastasiou:2015vya,Mistlberger:2018etf}, recently enhanced by the inclusion of the exact top-quark mass contributions up to three loops \cite{Czakon:2020vql}. The small theoretical uncertainty reached thanks to these higher-order corrections made secondary effects and productions mode relevant, as well as their associated uncertainty. Except for the lack of $\text{N}^3\text{LO}$ PDFs, the other sources of theoretical uncertainties all have sizes about $1\%$, and are given by the lack of full $c$- and $b$-quark mass inclusion, and the lack of an exact computation of NLO mixed QCD-Electroweak corrections.

Up to few years ago, mixed QCD-Electroweak corrections were known exactly only at LO \cite{Aglietti:2004nj,Degrassi:2004mx,Aglietti:2006yd}, while NLO results were first computed using a HEFT approach \cite{Anastasiou:2008tj}, and later in a purely massless case \cite{Anastasiou:2018adr}. Since NLO QCD corrections can produce large effects (up to $+100\%$, as seen in the purely HEFT-QCD case), the exact computation of the NLO corrections to Higgs boson gluon fusion is nowadays mandatory for providing the most reliable theoretical prediction.

In this contribution, based on \cite{Bonetti:2020hqh,Bonetti:2022lrk}, we provide the last missing ingredients for the full evaluation of NLO mixed QCD-Electroweak corrections to Higgs plus one jet production from proton collisions at the LHC.

\section{Processes analysis}
\label{sec:process}

\begin{figure}
\centering
	\subfloat[]{\includegraphics[width=0.20\textwidth]{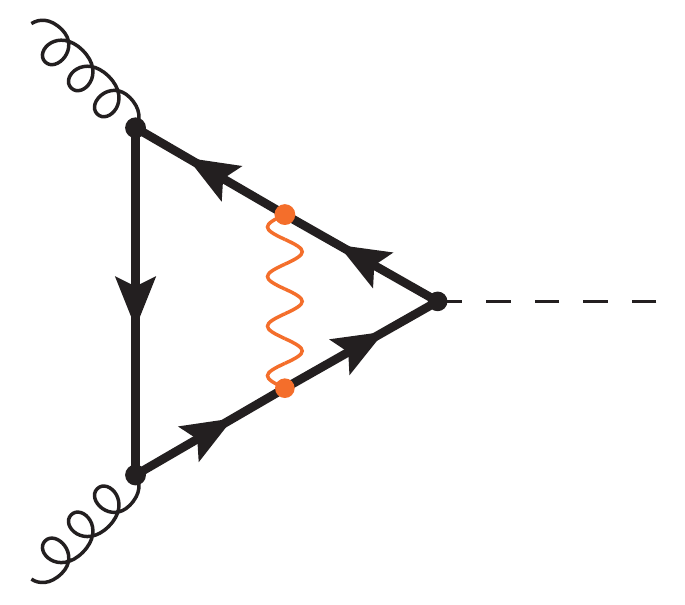}\label{fig:QCD-EW_modesA}}
	\qquad\qquad
	\subfloat[]{\includegraphics[width=0.20\textwidth]{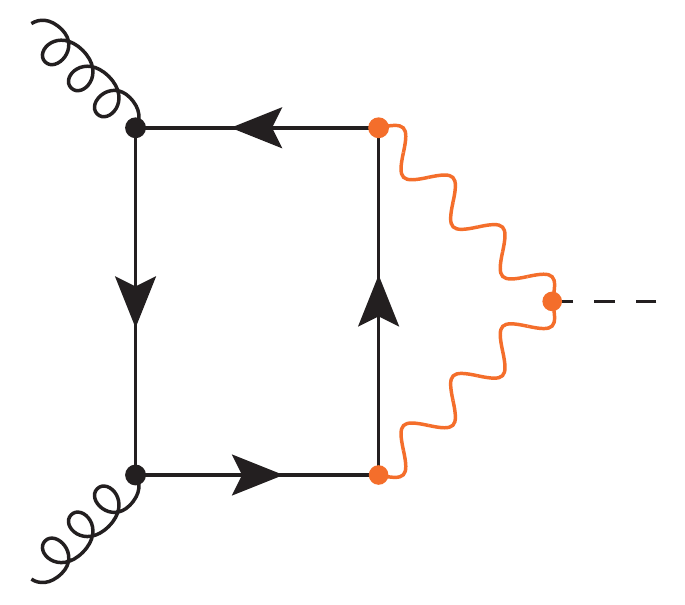}\label{fig:QCD-EW_modesB}}
	\caption{Different LO Electroweak corrections to Higgs boson production in gluon fusion.}
    \label{fig:QCD-EW_modes}
\end{figure}

Electroweak corrections to Higgs gluon fusion comprise two classes of diagrams, as depicted in Figure~\ref{fig:QCD-EW_modes}. On the one hand Electroweak bosons can occur inside the top-quark loop that connects the gluon and the Higgs (crf. Figure~\ref{fig:QCD-EW_modesA}); this kind of contributions are fairly small, producing corrections of the order of $0.5\%$ of the LO QCD cross section. On the other hand, Electroweak bosons generated by the quark loop can merge to produce the Higgs boson (crf. Figure~\ref{fig:QCD-EW_modesB}); such processes are dominated by light quarks, increase the total cross section of about $5\%$ with respect to the LO QCD one, and are our object of study.

\begin{figure}
\centering
	\subfloat[]{\includegraphics[width=0.20\textwidth]{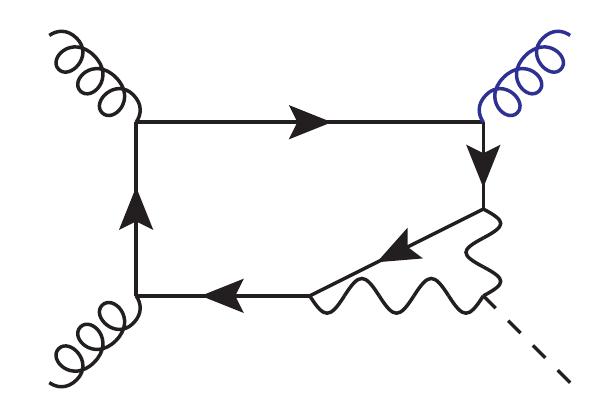}\label{fig:2loopCLOSED}}
	\qquad\qquad
	\subfloat[]{\includegraphics[width=0.20\textwidth]{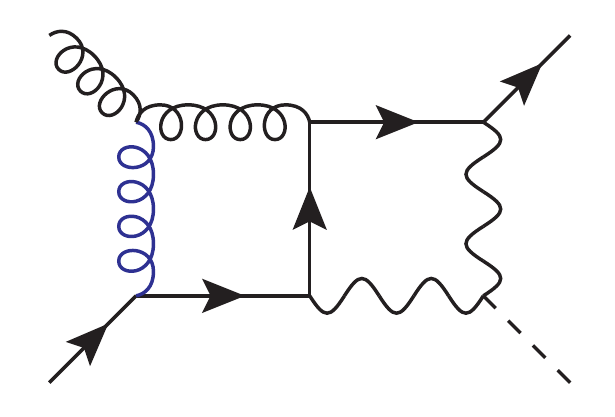}\label{fig:2loopOPEN}}
	\caption{Prototypical diagrams for $ggHg$ (closed fermion loop, left) and $q\overline{q}Hg$ (open fermion line, right).}
    \label{fig:2loop}
\end{figure}

Since we are interested in Higgs production starting from protons, both gluon-initiated and quark-initiated processes must be considered. In \cite{Aglietti:2004nj,Degrassi:2004mx,Aglietti:2006yd,Bonetti:2016brm,Bonetti:2017ovy} the two- and three-loop corrections to $ggH$ have been evaluated analytically with full dependence on the Electroweak and Higgs boson masses, and the same has been done at one-loop for the $q\overline{q}Hg$, $q\overline{q}Hgg$, and $q\overline{q}Hq\overline{q}$ in \cite{Becchetti:2020wof}. This leaves the two-loop corrections to $ggHg$ and to $q\overline{q}Hg$ to be evaluated (cfr. Figure~\ref{fig:2loop}).

The first difference with respect to the pure QCD case is given by chiral couplings stemming from vertices connecting quarks and Electroweak bosons. In case of closed fermion loops (as depicted in Figure~\ref{fig:2loopCLOSED}) terms containing a single $\gamma_5$ vanish summing over complete generations of massless quarks considering Standard Model couplings. The neat effect of such couplings is then reduced to an overall rescaling of the coupling constants, while the tensor structure of the amplitude is the same as for the pure QCD case
\begin{align}
	\label{eq:1}
    \mathcal{M} &= f^{c_1c_2c_3} \epsilon_{\lambda_1}^\mu(\mathbf{p}_1) \epsilon_{\lambda_2}^\nu(\mathbf{p}_2) \epsilon_{\lambda_3}^\rho(\mathbf{p}_3) \left[ \mathcal{F}_1 g_{\mu\nu}p_{2\rho} + \mathcal{F}_2 g_{\mu\rho} p_{1\nu} + \mathcal{F}_3 g_{\nu\rho} p_{3\mu} + \mathcal{F}_4 p_{3\mu} p_{1\nu} p_{2\rho} \right] ,
\end{align}
\begin{align}
    \mathcal{F}_{1\dots4} &= 4 F_{1\dots4,m_W} + \frac{2}{\cos^4 \theta_W}\left( \frac{5}{4} - \frac{7}{3}\sin^2\theta_W + \frac{22}{9}\sin^4\theta_W \right) F_{1\dots4,m_Z} \,.
\end{align}
A similar result is valid for open quark lines (cfr. Figure~\ref{fig:2loopOPEN}), although in this case a scheme conserving the anticommutativity of $\gamma_5$ is employed to move $\gamma_5$ to touch an external polarized spinor and convert into the corresponding eigenvalue. In this way we still obtain the same tensor structure as in QCD, but with a rescaling of the coupling depending on the chirality of the quark line
\begin{align}
    \mathcal{M}^{\text{open}}_L &= T^{c_3}_{i_1i_2} \left( \frac{2}{\cos^4 \theta_W} Q_q^2 \sin^4 \theta_W \right) \overline{v}_{s_1}(\mathbf{p}_1) \mathbb{P}_L \left[ \tau_{1,\mu} A_{1,m_Z}^{\textup{open}} + \tau_{2,\mu} A_{2,m_Z}^{\textup{open}} \right] u_{s_2}(\mathbf{p}_2) \epsilon^{\lambda_3}_\mu(\mathbf{p}_3)  \,,
\end{align}
\begin{align}
    \begin{aligned}
    \mathcal{M}^{\text{open}}_R &= T^{c_3}_{i_1i_2} \overline{v}_{s_1}(\mathbf{p}_1) \left\{\mathbb{P}_R \left[ \tau_{1,\mu} A_{1,m_W}^{\textup{open}} + \tau_{2,\mu} A_{2,m_W}^{\textup{open}} \right]  
		+\right.\\&\left.
		+ \frac{2}{\cos^4 \theta_W}\left( T_q - Q_q \sin^2 \theta_W \right)^2 \mathbb{P}_R \left[ \tau_{1,\mu} A_{1,m_Z}^{\textup{open}} + \tau_{2,\mu} A_{2,m_Z}^{\textup{open}} \right]\right\} u_{s_2}(\mathbf{p}_2) \epsilon^{\lambda_3}_\mu(\mathbf{p}_3)  \,,
	\end{aligned}
\end{align}
\begin{align}
	\label{eq:6}
    \begin{aligned}
	\mathcal{M}^{\text{closed}} & = T^{c_3}_{i_1i_2} \frac{1}{2} \overline{v}_{s_1}(\mathbf{p}_1) \left\{ 4 \left[ \tau_{1,\mu} F_{1,m_W}^{\textup{closed}} + \tau_{2,\mu} F_{2,m_W}^{\textup{closed}} \right] + \frac{2}{\cos^4 \theta_W}  \times\right.\\&\left. 
	\times\left( \frac{5}{4}-\frac{7}{3}\sin^2 \theta_W+\frac{22}{9}\sin^4 \theta_W \right) \left[ \tau_{1,\mu} F_{1,m_Z}^{\textup{closed}} + \tau_{2,\mu} F_{2,m_Z}^{\textup{closed}} \right] \right\} u_{s_2}(\mathbf{p}_2) \epsilon^{\lambda_3}_\mu(\mathbf{p}_3)  \,,
	\end{aligned}
\end{align}
\begin{align}
    \tau_{1,\mu} = \slashed{p}_3 p_{2\mu} - p_2 \cdot p_3 \gamma_\mu  \,,\qquad\qquad\tau_{2,\mu} = \slashed{p}_3 p_{1\mu} - p_1 \cdot p_3 \gamma_\mu \,.
\end{align}

\section{Evaluation of the form factors}
\label{sec:FFs}

\begin{figure}
\centering
	\subfloat[T1: $\FINB{1,1,1,1,0,1,1,1,0}$]{{\includegraphics[width=0.30\textwidth]{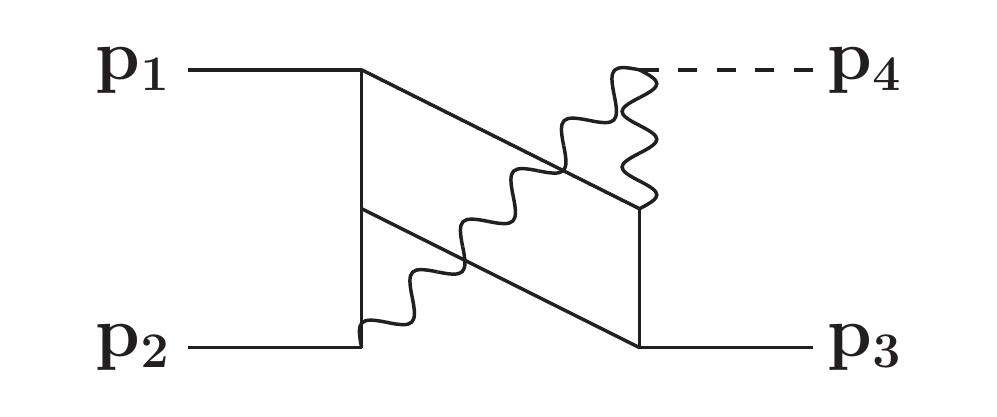}}\label{fig:T1}}
	\qquad
	\subfloat[T2: $\FINA{1,1,1,1,1,1,1,0,0}$]{{\includegraphics[width=0.30\textwidth]{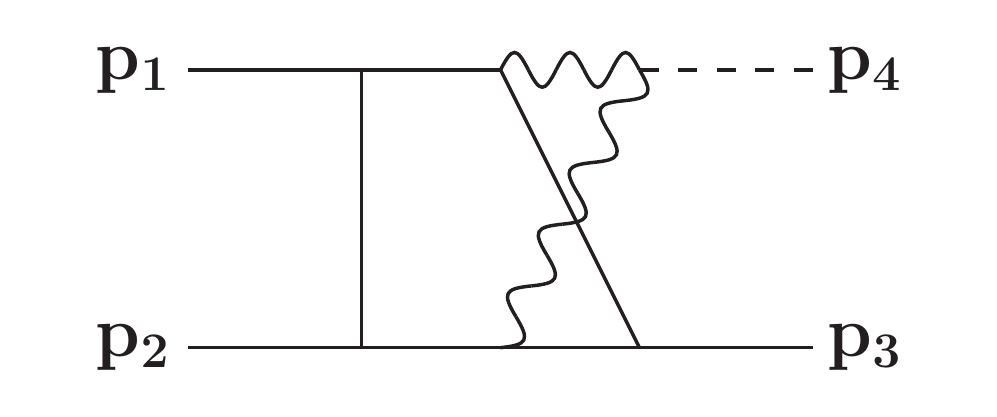}}\label{fig:T2}}
	\\
	\subfloat[T3: $\FIPL{0,1,1,1,1,0,1,1,1}$]{{\includegraphics[width=0.30\textwidth]{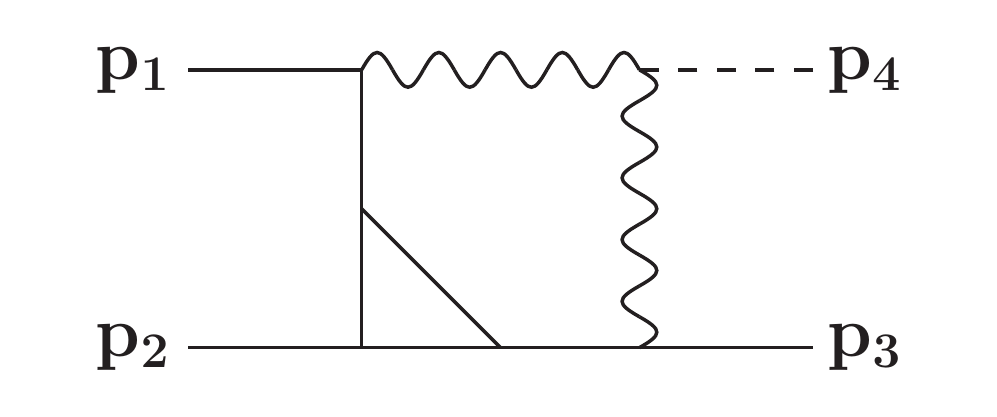}}\label{fig:T3}}
	\qquad
	\subfloat[T4: $\FIPL{1,1,1,1,0,0,1,1,1}$]{{\includegraphics[width=0.30\textwidth]{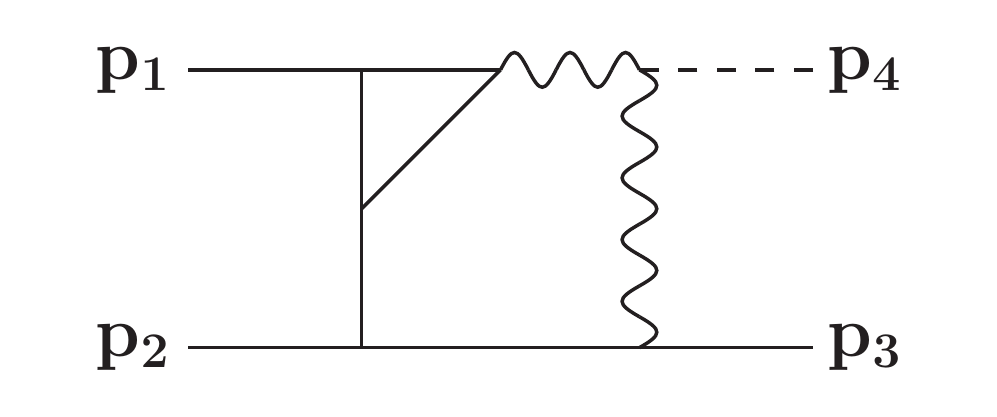}}\label{fig:T4}}
	\\
	\subfloat[T5: $\FIPL{0,1,1,1,1,1,1,1,0}$]{{\includegraphics[width=0.30\textwidth]{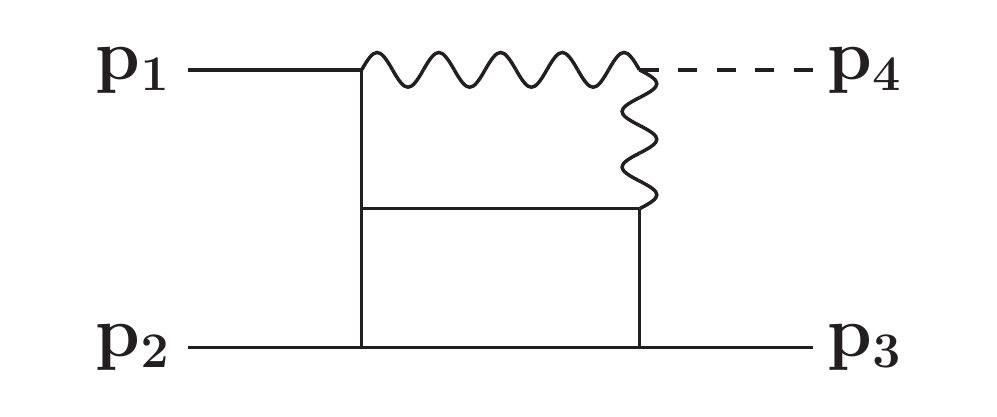}}\label{fig:T5}}
	\qquad
	\subfloat[T6: $\FIPL{1,1,1,1,1,1,1,0,0}$]{{\includegraphics[width=0.30\textwidth]{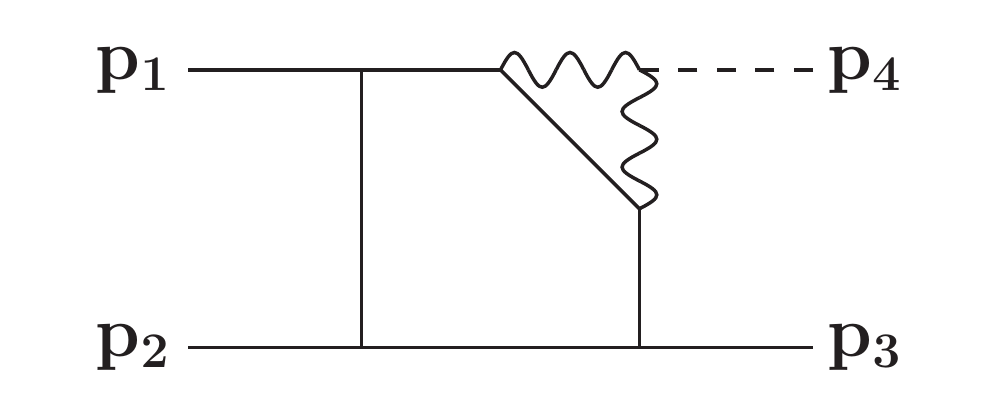}}\label{fig:T6}}	
	\caption{The six top sectors appearing in the amplitude. Straight (wavy) lines denote massless (massive) propagators. The dashed line indicates the Higgs boson.}
    \label{fig:topolo}
\end{figure}

We construct and apply projectors in order to extract the single form factors $F$ of Eqs.~\eqref{eq:1} to \eqref{eq:6} in terms of two-loop Feynman integrals. Such integrals are (sub-)graphs of the six top sectors depicted in Figure~\ref{fig:topolo} (T2 and T6 are enough to describe the integrals appearing in the $ggHg$ amplitude, while the other four sectors are necessary to cover the integrals coming from the $q\overline{q}Hg$ process). We use the computer programs \texttt{Reduze} \cite{vonManteuffel:2012np} and \texttt{Kira} \cite{Maierhoefer:2017hyi} to reduce these Feynman integrals to a basis of Master Integrals (61 for the $ggHg$ case, plus other 30 for the $q\overline{q}Hg$ case). The presence of the eight different square roots
\begin{align}
	\label{eq:roots}
	\begin{aligned}
		r       &= m_h^2 \sqrt{1-4 m_V^2/m_h^2} \,,\qquad&&
        r_{ust} &= \sqrt{s^2 u^2 +2su(t-s)m_V^2+(s+t)^2m_V^4}    \,,\\
        r_t     &= \sqrt{r^2-4 m_V^2 su/t}      \,,\qquad&&
        r_{sut} &= \sqrt{s^2 u^2+2su(t-u)m_V^2+(t+u)^2m_V^4}     \,,\\
        r_u     &= \sqrt{r^2-4 m_V^2 st/u}      \,,\qquad&&
        r_{stu} &= \sqrt{s^2 t^2 + 2st(u-t)m_V^2+(t+u)^2m_V^4}   \,,\\
        r_{tu}  &= \sqrt{1-4 m_V^2 /(t+u)}      \,,\qquad&&
        r_{uts} &= \sqrt{u^2 t^2 + 2ut(s-t)m_V^2+(s+t)^2m_V^4}   \,,
    \end{aligned}
\end{align}
which, to our knowledge, are not simultaneously rationalizable, make the usage of differential equations ineffective to solve the master integrals. We then revert to direct integration over Feynman--Schwinger parameters. We find that all our master integrals are linearly reducible \cite{Panzer:2014caa}, i.e.\ there exists an order of integration such that each integral returns a hyperlogartihmic expression in terms of the next integration variable. This has the advantages that on the one hand we do not need to rationalize any square root (since they appear in the kinematics) and that on the other hand our result will still be written in terms of Goncharov polylogarithms (or GPLs), explicitly containing the square roots from Eq.~\eqref{eq:roots}.

It is important to notice that, despite the fact that two-loop integrals might contain poles up to $\epsilon^{-4}$, we expect the $ggHg$ amplitude to be finite and the $q\overline{q}Hg$ one to have at most $\epsilon^{-2}$ singularities. To exploit this simpler pole structure we construct a (quasi-)finite basis of master integrals, removing UV- and IR-divergent terms. UV singularities are cured by raising the powwer of massive propagators until we reach a negative degree of divergence, while IR singularities are tackled by moving from four to six dimensions \cite{Tarasov:1996br,Lee:2009dh}
\begin{align}
	\mathcal{I}^{D+2}(a_1,\dots,a_7) = \frac{16}{s t u (D-4) (D-3)} \int \td^D k_1 \td^D k_2 \frac{G(k_1,k_2,p_1,p_2,p_3)}{\mathcal{D}_1^{a_1} \dots \mathcal{D}_7^{a_7}}
\end{align}
where the Gram determinant $G$ counters soft and collinear divergencies.

We compute the analytic $\epsilon$-expansions of the master integrals using {\HyperInt} \cite{Panzer:2014caa} and insert them in the amplitudes, obtaining a linear combination of transcendental functions with algebraic coefficients. UV divergences and IR poles present in the $q\overline{q}Hg$ amplitude are removed by renormalization of $\alpha_S$ and subtraction of the Catani operator (see \cite{Catani:1998bh} for further details).

We proceed now by simplifying the finite remainders. As a first step we use the \texttt{Mathematica} package \texttt{MultivariateApart} \cite{Heller:2021qkz} to partial-fraction our expressions without introducing spurious poles in the kinematics. We reduce then the set of algebraic prefactors to a basis of algebraic functions, collecting the new transcendental expressions and numerically checking for null terms. We scan the remaining transcendental expressions with PSLQ, finding further linear relations that we employ to write transcendental expressions of higher weights in terms of simpler objects. We further simplify the $ggHg$ amplitude by rewriting GPLs up to weight 3 in terms of $\Li{2}$, $\Li{1,1}$, and $\log$s, making analytic continuation explicit in the physical region of the process. At last, we combine the form factors to obtain the helicity amplitudes. For the $ggHg$ case they read
\begin{align}
    \mathcal{A}^{ggHg}_{+++} &= \frac{m_h^2}{\sqrt{2}\langle12\rangle\langle23\rangle\langle31\rangle} \frac{s u}{m_h^2} \left( \mathcal{F}_1 + \frac{t}{u}\mathcal{F}_2 + \frac{t}{s}\mathcal{F}_3 + \frac{t}{2}\mathcal{F}_4 \right) ,
\\
    \mathcal{A}^{ggHg}_{++-} &= \frac{[12]^3}{\sqrt{2}m_h^2[13][23]} \frac{u m_h^2}{s} \left( \mathcal{F}_1 + \frac{t}{2}\mathcal{F}_4 \right) ,
\end{align}
while for the $q\overline{q}Hg$ process we have
\begin{align}
    \begin{aligned}
        \mathcal{A}^{q\overline{q}Hg}_{RL+} &= \left\{\frac{1}{2}\left[ 4 \mathcal{F}_{1,m_W}^{\textup{closed}} 
	+ \frac{2}{\cos^4 \theta_W}\left( \frac{5}{4}-\frac{7}{3}\sin^2 \theta_W+\frac{22}{9}\sin^4 \theta_W \right) \mathcal{F}_{1,m_Z}^{\textup{closed}} \right]
			+\right.\\&\qquad\left.+
			\left[ \mathcal{F}_{1,m_W}^{\textup{open}} + \frac{2}{\cos^4 \theta_W}\left( T_q - Q_q \sin^2 \theta_W \right)^2 \mathcal{F}_{1,m_Z}^{\textup{open}} \right] \right\} \frac{s}{\sqrt{2}}\frac{[23]^2}{[12]} ,
	\end{aligned}
\end{align}
\begin{align}
    \begin{aligned}
    \mathcal{A}^{q\overline{q}Hg}_{LR+} &= \left\{\frac{1}{2}\left[ 4 \mathcal{F}_{2,m_W}^{\textup{closed}} 
	+ \frac{2}{\cos^4 \theta_W}\left( \frac{5}{4}-\frac{7}{3}\sin^2 \theta_W+\frac{22}{9}\sin^4 \theta_W \right) \mathcal{F}_{2,m_Z}^{\textup{closed}} \right]
			+\right.\\&\qquad\left.+
			\left[ \frac{2}{\cos^4 \theta_W} Q_q^2 \sin^4 \theta_W \mathcal{F}_{2,m_Z}^{\textup{open}} \right]  \right\} \frac{s}{\sqrt{2}}\frac{[13]^2}{[12]} .
	\end{aligned}
\end{align}

\section{Conclusions}
\label{sec:conlcusions}

We evaluated the missing ingredients to the NLO mixed QCD-Electroweak corrections to Higgs production at the LHC. We obtained analytic expressions for the helicity amplitudes with full dependence over the Electroweak and Higgs boson masses, written in terms of GLPs, $\Li{}$, and $\log$ functions. With all the analytic expressions for the amplitudes readily available, it is possible to compute observables, starting with the total cross section for Higgs plus jet production including both gluon- and quark-initiated channels, refining the results presented in \cite{Becchetti:2020wof}.

So far we have considered light-quark contributions, although the top quark is a necessary ingredient to fully include the third family of fermions in the picture. The fact that the top quark is massive will introduce extra steps on both the tensorial manipulation of the amplitude, which will now require the consistent usage of a so-called \emph{$\gamma_5$ scheme} to generalize the notion of chirality from 4 to $D$ dimensions, and on the reduction and evaluation of master integrals, introducing much more demanding steps for finding relations among the integrals and for evaluating them, probably requiring series expansions in the kinematics. 

~

\textbf{Acknowledgments.}~These contributions are based on \cite{Bonetti:2020hqh,Bonetti:2022lrk}, by Erik Panzer, Vladimir A.\ Smirnov, and Lorenzo Tancredi. M.B.\ is supported by the Deutsche Forschungsgemeinschaft (DFG) under grant no.\ 396021762 - TRR 257.

\bibliographystyle{JHEP}
\bibliography{Biblio}

\end{document}